\documentclass[1p,12pt]{elsarticle}

\usepackage{amssymb}
\usepackage{amsmath}
\usepackage{here} 
\usepackage{cleveref} 
\usepackage{dcolumn} 
\usepackage{natbib}
\usepackage{multirow}
\usepackage{tabularx}
\newcommand\T{\rule{0pt}{2.6ex}}       
\newcommand\B{\rule[-1.2ex]{0pt}{0pt}} 

\def\url#1{}
\newcolumntype{Y}{>{\centering\arraybackslash}X}

\journal{Advances in Quantum Chemistry}

\begin{document}

\begin{frontmatter}



\title{Accurate Born-Oppenheimer potentials for excited $\Sigma^{+}$ states of the hydrogen molecule}


\author{Micha{\l} Si{\l}kowski\corref{cor}}
\ead{michal.silkowski@fuw.edu.pl}
\author{Magdalena Zientkiewicz}
\author{Krzysztof Pachucki}

\address{Faculty of Physics, University of Warsaw, Pasteura 5, 02-093 Warsaw, Poland}

\cortext[cor]{Corresponding author}

\begin{abstract}
We report on highly accurate calculations of Born-Oppenheimer potentials for excited $n\,\Sigma^{+}$ states of the hydrogen
molecule for all possible combinations of singlet/triplet and gerade/ungerade symmetries up to $n=7$. A~relative
accuracy of $10^{-10}$ (0.00002 cm$^{-1}$) or better is achieved for all the internuclear distances and all the excited
states under consideration -- an~improvement with respect to the best results available in the literature by at least 6 orders of magnitude.
Presented variational calculations rely on efficient evaluation of molecular integrals with the explicitly correlated
exponential basis in arbitrary precision arithmetics.
\end{abstract}


\end{frontmatter}



\section{Introduction}

\emph{Ab initio} Born-Oppenheimer potentials have proven to be invaluable for the interpretation of spectroscopic data.
Currently, transitions between rovibrational levels of $X\,^1\Sigma^{+}_g$ ground state for hydrogen molecule isotopologues are a
subject of exhaustive investigations both from experimental \cite{ubachssm,niu,laidt} and theoretical
\cite{naptrel,naqed} standpoints, reaching the accuracy  of $10^{-4}$ cm$^{-1}$. On the contrary, electronically excited
states of the hydrogen molecule are much less studied, especially by theoretical methods. So far, calculations
of Born-Oppenheimer (BO) potentials for excited $\Sigma^{+}$ states of H$_2$ were performed with full CI
method~\cite{corongiugs,corongiuus,corongiut}, variational calculations with generalized GTOs~\cite{detmer}, variational
calculations with explicitly correlated gaussians (ECG)~\cite{komasaecg,komasaef} and more recently with free-complement
local-Schr{\"o}dinger-equation method (FC-LSE) \cite{nakatsuji}. The most accurate to date
are variational calculations in an explicitly correlated exponential basis by Wolniewicz and co-workers
\cite{woldressler,orlikowski,staszewska1999,staszewska2002,staszewska2003} and by Komasa \emph{et al.}
\cite{komasaecg,komasaef} which, remarkably have remained unsurpassed for the last 20 years.

Among the plethora of H$_2$ excited states, of particular importance and long-standing interest is the $B\,1^{1}\Sigma^{+}_u$ state, the lowest of ${}^1\Sigma^{+}_u$ symmetry.
It is~easily accessible for spectroscopic measurements, through the allowed electric dipole transition from the
$X\,1^{1}\Sigma^{+}_g$ ground state, therefore, it can be measured very accurately and serve as a calibration marker
at XUV frequencies.
Likewise, energies of states such as $EF\,2^1\Sigma^{+}_g$, $GK\,3^1\Sigma^{+}_g$, $B\,1^1\Sigma^{+}_u$ and
$B'\,2^1\Sigma^{+}_u$ are of great spectroscopic interest, because they comprise intermediate states in multiphoton processes, which are
crucial for investigations of doubly excited states, recognized as resonant states involved in the dynamics of preionization
and predissociation \cite{lai,gk,dynamics}. 

Moreover, BO potentials comprise an essential input for Multichannel Quantum Defect theory analyses of
Rydberg states, where an accuracy of BO energies better than $0.001\,\mathrm{cm}^{-1}$ is required
\cite{mqdt,ortopara}, especially for the highly excited states,
despite their minor spectroscopic relevance due to their experimental inaccessibility from the ground state.
Accurate potentials are also important for studying  \emph{gerade}/\emph{ungerade} 
mixing effects \cite{merkt} beyond the BO approximation.
Ultimately, accurate BO potentials may serve as a useful benchmark for less precise computational methods, electron-ion scattering experiments and numerous other applications~\cite{takahashi,glassm,edwards}.
We~note that many previous calculations often come without estimation of uncertainties,
making definitive comparisons rather difficult.

\section{Method}

The variational BO potentials reported in this work are obtained with the use of explicitly correlated exponential
functions~\cite{kw} of the form,
\begin{equation}
	\Psi_{\Sigma} = \sum_{\{n\}} c_{\{n\}} \left(1 \pm P_{AB} \right) \left(1 \pm P_{12} \right) \Phi_{\{n\}},
\end{equation}
where $P_{AB}$ permutes the nuclei $A$ and $B$, $P_{12}$ interchanges the two electrons and appropriate $\pm$ signs are chosen
to fulfil the symmetry criteria for \textit{gerade}/\emph{ungerade} and \textit{singlet}/\textit{triplet} states. Linear coefficients
$c_{\{n\}}$ form an eigenvector, which is a solution of a secular equation. Trial wavefunctions $\Phi_{\{n\}}$ are given by,
\begin{eqnarray}
\Phi_{\{n\}} &=&
	e^{-y\,\eta_1 -x\,\eta_2 -u\,\xi_1 -w\,\xi_2}
	r_{12}^{n_0} \, \eta_1^{n_1} \, \eta_2^{n_2} \, \xi_1^{n_3} \, \xi_2^{n_4},
\end{eqnarray}
where $\eta_i$ and $\xi_i$ are proportional to confocal elliptic coordinates and are given by $\eta_i=r_{iA}-r_{iB}$,
$\xi_i=r_{iA}+r_{iB}$, $y,x,u$ and $w$ are real, nonlinear parameters subject to variational minimization. The index $\{n\}$ denotes $\{n\} = (n_0,n_1,n_2,n_3,n_4)$ and enumerates all powers of coordinates
allowed by symmetry restrictions, with an additional constraint,
\begin{equation}
	\sum_{j=0}^4 n_j \le \Omega,
\end{equation}
which determines the total size of the basis, once $\Omega$ is set.
Analogously to the Hylleraas basis, a trial wavefunction is uniquely characterized by the set of $\{n\}$ and nonlinear
parameters, which are common to all trial functions within a given sector.

In this work we use two sectors, each with its own nonlinear parameters, and with no additional symmetry restrictions.
With $\Omega$ determining the total size of the first sector, the second sector's principal number is chosen as $\Omega - 2$, a
heurestic optimum for most cases.
In previous works \cite{h2bo,gu} concerning ground $X\,{}^{1}\Sigma^{+}_g$ or $b\,{}^3 \Sigma^{+}_u$ states the James-Coolidge ($x=y=0$) or
generalized Heitler-London ($x=-y, u=w$) was utilized.
The reasoning behind the more general basis used in this work stems from a fairly universal trend of wavefunctions being
composed predominantly of ($1s,n'l$) configurations with high values of $n'$ and $l$ \cite{corongiugs,corongiuus,corongiut,mulliken}, where in the
language of atomic orbitals, one electron is highly excited.
This implies substantial asymmetry, $x\ne y$ and $u\ne w$, and demands a more flexible basis to accurately represent the
wavefunction. In addition, the contribution of ionic configuration ($1s',1s'$) grows more significant not only for
singlet \cite{staszewska2002,corongiugs,corongiuus} but also for triplet symmetry~\cite{corongiut}, and there are states in the ${}^1\Sigma^{+}$ manifolds where it is essential~\cite{kolosb} for significantly broad
ranges of internuclear distances.

Thirdly, in contrast to the $X\, {1}\Sigma^{+}_g$ and $b\, ^3 \Sigma^{+}_u$ states, BO potentials for higher $n$ exhibit much
more complicated structure -- a result of mutual avoided crossings and resonant interaction with the H${}^{+}$H${}^{-}$
configurations, often resulting in double minima or bumps, rather than a smooth, single minimum Morse-like curve. This, in
turn, implies the presence of strong configuration mixing in some regions of $R$, unlike the case of well-separated $X\, ^{1}\Sigma^{+}_g$ and $b\, ^3 \Sigma^{+}_u$ states.

Justification for introducing a second sector into the basis set follows from the presence of distinct scales of
motion, namely, around the nuclei and another for larger distances \cite{sims, drake}.
Many previous calculations utilize specific structure of the basis, tailored for a single state or symmetry, often
with \emph{a priori} knowledge about the chemical bond character or atomic orbitals approximation. 
Here, we emphasize the universality of our approach -- nonlinear parameters for both sectors are thoroughly optimized in
an unrestricted Kolos-Wolniewicz basis \cite{kolosb,rychlewskibook} separately for each state and every internuclear distance. This basis is
capable of accurately representing ionic and covalent structure with correct long-range asymptotics and short-range
vicinities of interparticle cusps, as well as complicated radial nodes, which are expected in highly excited states.

A crucial point of our computational method is that it relies on the fact that all the necessary matrix elements of the nonrelativistic Hamiltonian,
\begin{equation}
H =
-\frac{\nabla_1^2}{2}-\frac{\nabla_2^2}{2}-\frac{1}{r_{1A}}-\frac{1}{r_{1B}}
-\frac{1}{r_{2A}}-\frac{1}{r_{2B}} +\frac{1}{r_{12}}
+\frac{1}{r_{AB}},
\end{equation}
can be readily constructed as a linear combination of $f$-integrals,
which are given by differentiation with respect to the corresponding nonlinear parameter, with $r=r_{AB}$,
\begin{align}
f(r,n_0,n_1,n_2,n_3,n_4) =  \frac{(-1)^{n_0+n_1+n_2+n_3+n_4}}{n_0!\,n_1!\,n_2!\,n_3!\,n_4!}
\frac{\partial^{n_0}}{\partial w_1^{n_0}}\biggr|_{w_1 = 0}\,
\frac{\partial^{n_1}}{\partial y^{n_1}}\,
\frac{\partial^{n_2}}{\partial x^{n_2}}\,
\frac{\partial^{n_3}}{\partial u^{n_3}}\,
\frac{\partial^{n_4}}{\partial w^{n_4}}\,f(r)\,,
\label{03}
\end{align}
of $f(r)$, called the master integral \cite{kw},
\begin{equation}
f(r) = r\,\int \frac{d^3 r_1}{4\,\pi}\,\int \frac{d^3 r_2}{4\,\pi}\,
\frac{e^{ -w_1\,r_{12} - u\,(r_{1A} + r_{1B}) - w\,(r_{2A} + r_{2B})
                    - y\,(r_{1A} - r_{1B}) - x\,(r_{2A} - r_{2B})}}
{r_{12}\,r_{1A}\,r_{1B}\,r_{2A}\,r_{2B}}. \label{master}
\end{equation}
The master integral (\ref{master}) satisfies the differential equation \cite{rec2h2,lesiuk},
\begin{equation}
\biggl[\sigma_4\,\frac{d^2}{d\,r^2}\,r\,\frac{d^2}{d\,r^2} + \sigma_2\,\frac{d}{d\,r}\,r\,\frac{d}{d\,r}
+ \sigma_0\,r\biggr]\,f(r)  = F(r), \label{04}
\end{equation}
where
\begin{eqnarray}
\sigma_0 &=& w_1^2\,(u + w - x - y)\,(u - w + x - y)\,(u - w - x + y)\,(u + w + x + y) \nonumber \\&&
        + 16\,(w\,x - u\,y)\,(u\,x - w\,y)\,(u\,w - x\,y),\nonumber \\
\sigma_2 &=& w_1^4-2\,w_1^2\,(u^2+w^2+x^2+y^2)+16\,u\,w\,x\,y,\nonumber \\
\sigma_4 &=& w_1^2\,,\label{05}
\end{eqnarray}
and $F(r)$ is an inhomogeneous term, which involves exponential functions, exponential integral functions Ei, and the natural
logarithm, with nonlinear parameters as their arguments, see Eq.\,(7) of Ref. \cite{kw} for explicit expression.

The differential equation (\ref{04}) entails that the master integral $f(r)$ and all its derivatives with respect to nonlinear
parameters have a Taylor-like expansion in $r$,
\begin{equation}
f(r) = \sum_{n=1}^\infty \bigl[f^{(1)}_n\,(\ln (r) +\gamma_{\rm E}) + f^{(2)}_n\bigr]\,r^n\,, \label{10}
\end{equation}
where $\gamma_{\rm E}$ is the Euler-Mascheroni constant. Derivation of efficient recursion relations for evaluation of
coefficients $f^{(1)}_n$ and $f^{(2)}_n$ constitutes an essential step towards evaluating $f$-integrals with
arbitrary $y,x,u$ and $w$ nonlinear parameters and arbitrary precision, see Ref. \cite{kw} for more details.

Once all matrix elements are constructed in terms of $f$-integrals, the electronic energy is found as the $n$th lowest
eigenvalue by refining an initial guess with inverse iteration method. This step is the most time consuming part of
our computational method, because it scales
as $\mathcal{O}(N^3)$, where $N$ is the total size of the basis.
Recently, interest in Aasen's algorithm \cite{aasen} for indefinite matrix factorization has been revived due to a
novel communication-avoiding variant \cite{block}, suitable for efficient parallelization on modern computer
architectures.

In our calculations a custom implementation of this algorithm inspired by the PLASMA
library~\cite{plasma} was utilized with custom vectorization of quad-double arithmetics based on Bailey's
quad-double precision algorithms~\cite{qd}.
Incorporation of both the communication-avoiding variant of Aasen's factorization and vectorized quad-double arithmetics
constitutes a very efficient implementation of dense matrix factorization -- an essential component of the inverse iteration method with nonorthogonal bases. Extension to octuple precision arithmetics has proven sufficient to contend with
near-singular matrices, ill-conditioning of which can be attributed to the presence of large powers of the $r_{12}$ coordinate in the basis.

\section{Results and discussion}

Reported BO energies are calculated for all possible symmetries of the $\Sigma^{+}$ manifold: $n^{1}\Sigma^{+}_g$, $n^{3}\Sigma^{+}_g$,
$n^{1}\Sigma^{+}_u$, and $n^{3}\Sigma^{+}_u$, with $n\le 7$ (26 states in total), with the exception of the ground
$X\,{}^{1}\Sigma^{+}_g$ and $b\,{}^3 \Sigma^{+}_u$ states, which were calculated elsewhere \cite{h2bo,gu} using special
cases of the KW basis. The results are obtained on a grid of internuclear distances resembling logarithmic spacing in
the range $R=0.7-20.0$ au (72 points per state in total). It is sampled with denser spacing of $0.1$ au in the region
of $R=4.0-6.0$ au, where potential curves demonstrate strong avoided crossing features and are subject to rapid changes as a function of $R$. 

\begin{figure}[htb]
\centering
\includegraphics[width=\columnwidth]{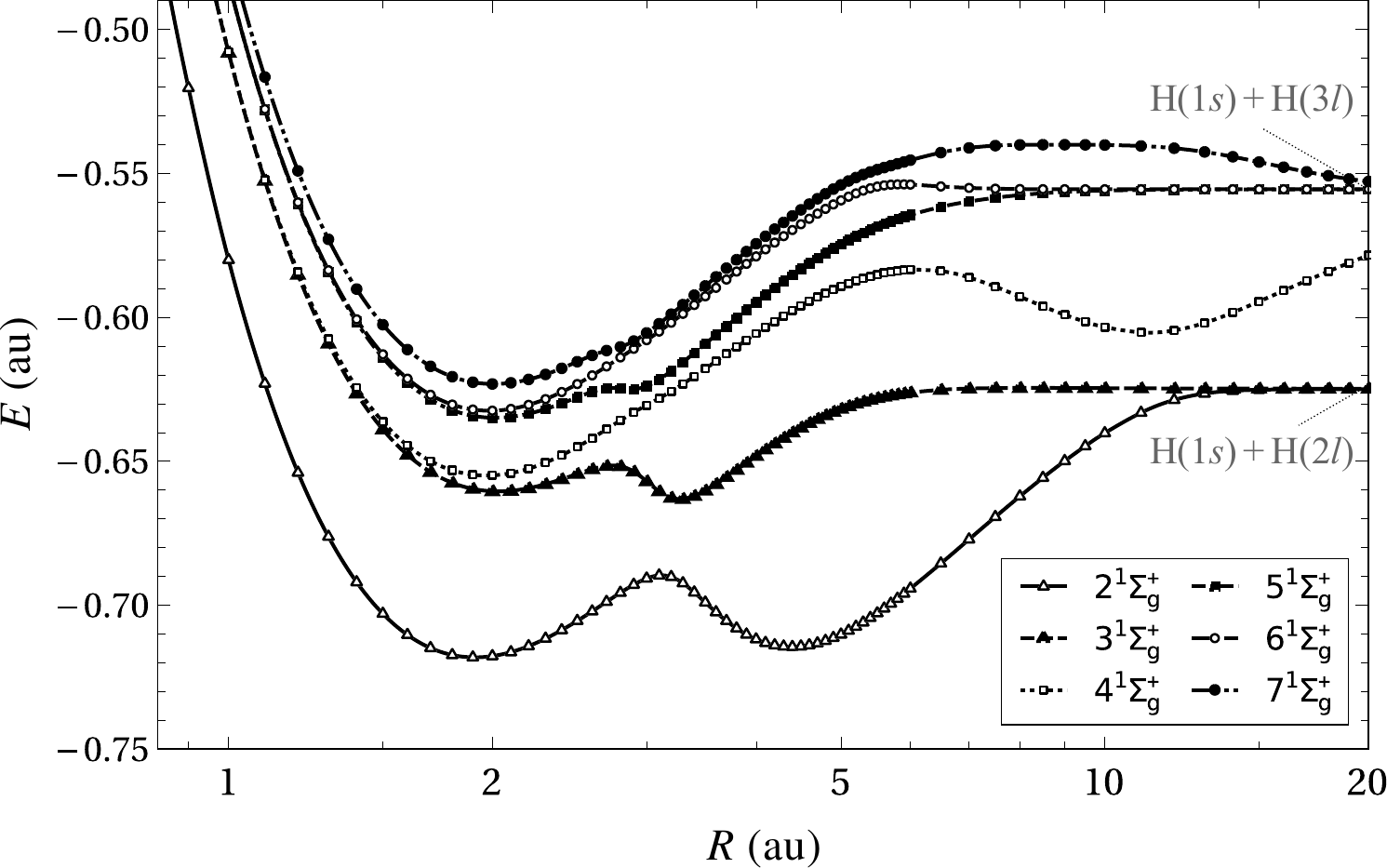}  
\caption{ BO potentials for $n^1\Sigma^{+}_g$ states, up to $n=7$. In this scale the $X\,1^1\Sigma^{+}_g$ ground state
is not visible. The $GK\,3^1\Sigma^{+}_g$ with $H\,4^1\Sigma^{+}_g$ and $P\,5^1\Sigma^{+}_g$ with $O\,6^1\Sigma^{+}_g$
switch character from united atom $3d\sigma$ and $4d\sigma$ configurations to $3s$ and $4s$, respectively, exhibiting
$<1\,\rm{cm}^{-1}$ splitting around $R=1.0$ au \cite{woldressler}.}
\label{sigs}
\end{figure}

Numerical results for BO energies obtained in this work, together with uncertainties originating purely from extrapolation to the complete basis set (CBS) limit are
presented in Tables S1 -- S26 in the Supplementary Material \cite{supp}. Additional quantities such as $dE/dR$, $-\langle \nabla_1^2 \rangle/2$ and $-\langle\nabla_1 \cdot \nabla_2\rangle/2$ along with $E$ itself are reported in raw format of the Supplementary Material \cite{supptxt}.
Unsurprisingly, convergence for triplet states is better than for singlet, because the requirement of antisymmetry of total
fermionic wavefunction generally results in less electron-electron correlation -- well-known phenomenon of \emph{correlation hole}.

\begin{table}[htb]
	\centering
	\footnotesize
	\caption{Exemplary values of optimal nonlinear parameters for states and internuclear distances with high ionic component.}
	\begin{tabular}{cc|cccc|cccc}
		\hline
		\hline
		\multirow{2}{*}{state} 		   &  \multirow{2}{*}{ $R$ }   & \multicolumn{4}{c|}{sector 1} &  \multicolumn{4}{c}{sector 2} \T \\
		\cline{3-10}
		     & & $ y$ & $x$ & $u$ & $w$ & $ y$ & $x$ & $u$ & $w$ \\
			 \hline
		$B\,1^1\Sigma^{+}_u$         & 10.0 &  0.047 & -0.447 & 0.338 & 0.468 & -0.949 & -1.263 & 1.052 & 1.226 \T \\
		$H\bar{H}\,4^1\Sigma^{+}_g$ & 12.0 &  0.000 & -0.446 & 0.263 & 0.506 & -0.807 & -0.807 & 1.068 & 1.068 \\
		$6^1\Sigma^{+}_u$           & 15.0 &  0.079 & -0.490 & 0.163 & 0.493 & -0.863 & -0.727 & 0.847 & 0.520 \\
		$7^1\Sigma^{+}_g$           & 20.0 & -0.357 & -0.858 & 0.276 & 0.692 & -0.428 &  0.164 & 0.445 & 0.146 \\
	\hline
	\hline
\end{tabular}
\label{tab:ionic}
\end{table}

\begin{figure}[h]
\centering
\includegraphics[width=\columnwidth]{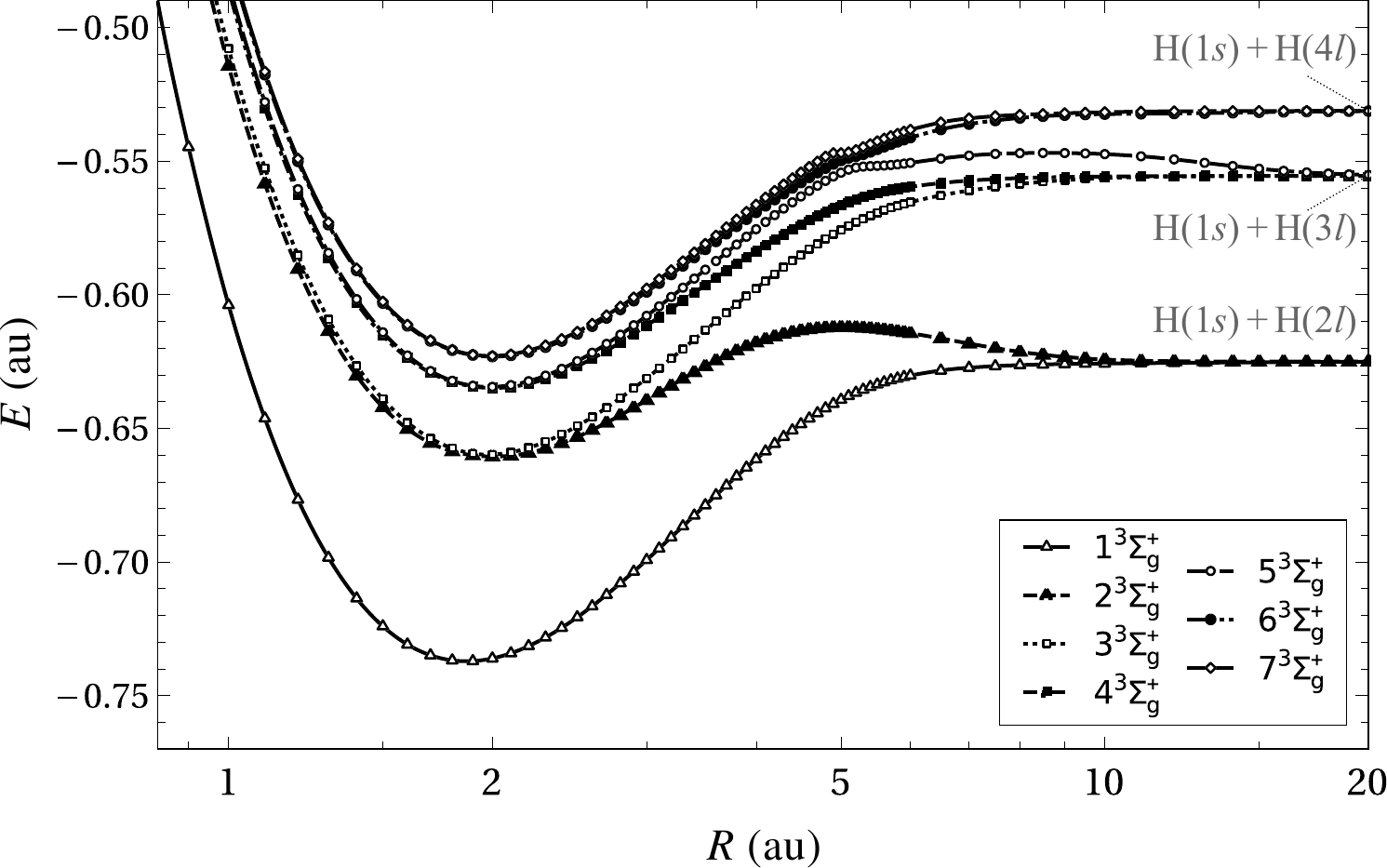}  
\caption{ BO potentials for $n^3\Sigma^{+}_g$ states, up to $n=7$. Pairs of states $2-3$, $4-5$ and $6-7$ all anticross
around the $R=2.0$ au minima, conjointly changing the character from united atom $3d\sigma,4d\sigma,5d\sigma$ to $3s,4s,5s$, respectively.}
\label{sigt}
\end{figure}

\begin{figure}[h]
\centering
\includegraphics[width=\columnwidth]{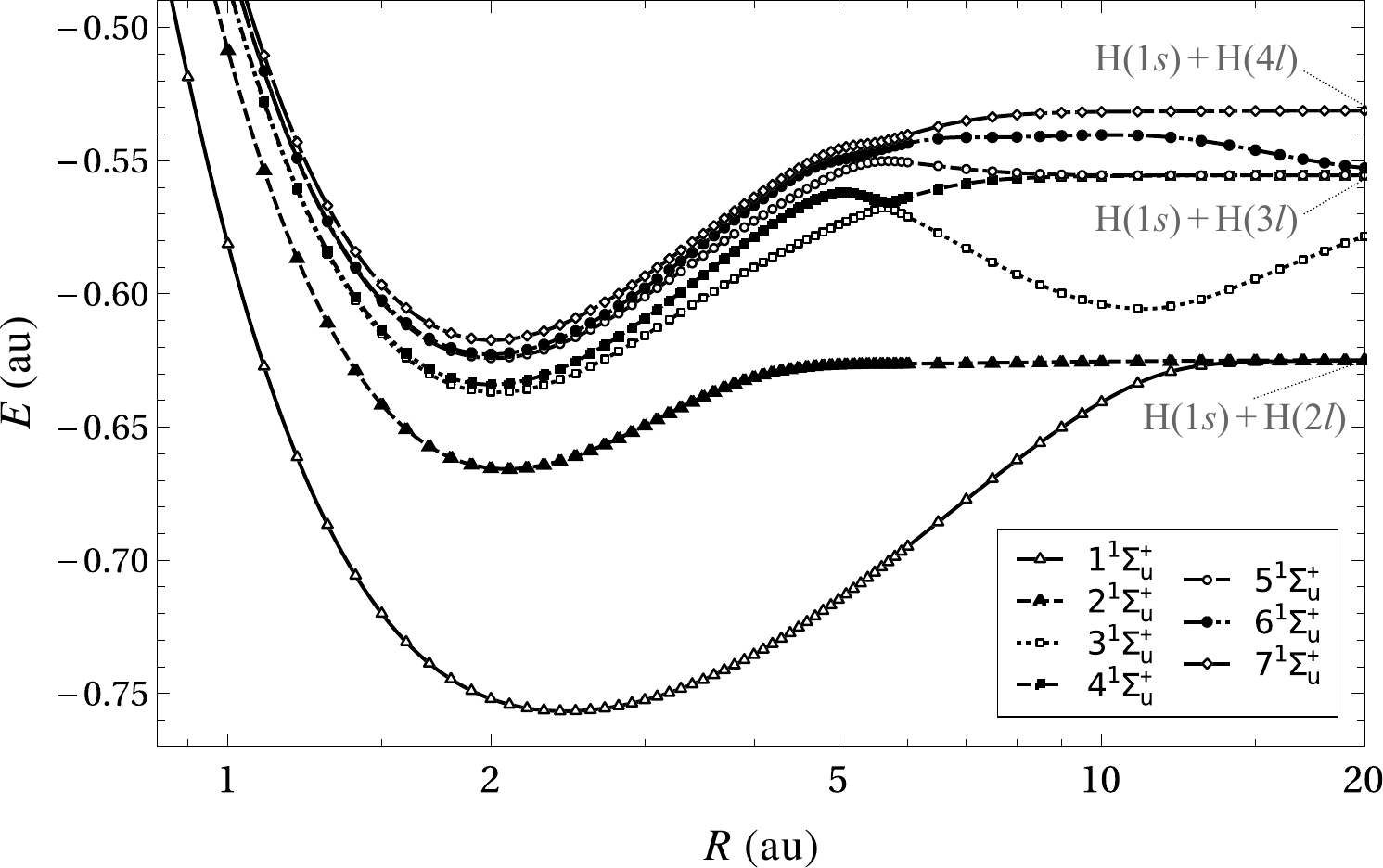}  
\caption{ BO potentials for $n^1\Sigma^{+}_u$ states, up to $n=7$.}
\label{sius}
\end{figure}

\begin{figure}[h]
\centering
\includegraphics[width=\columnwidth]{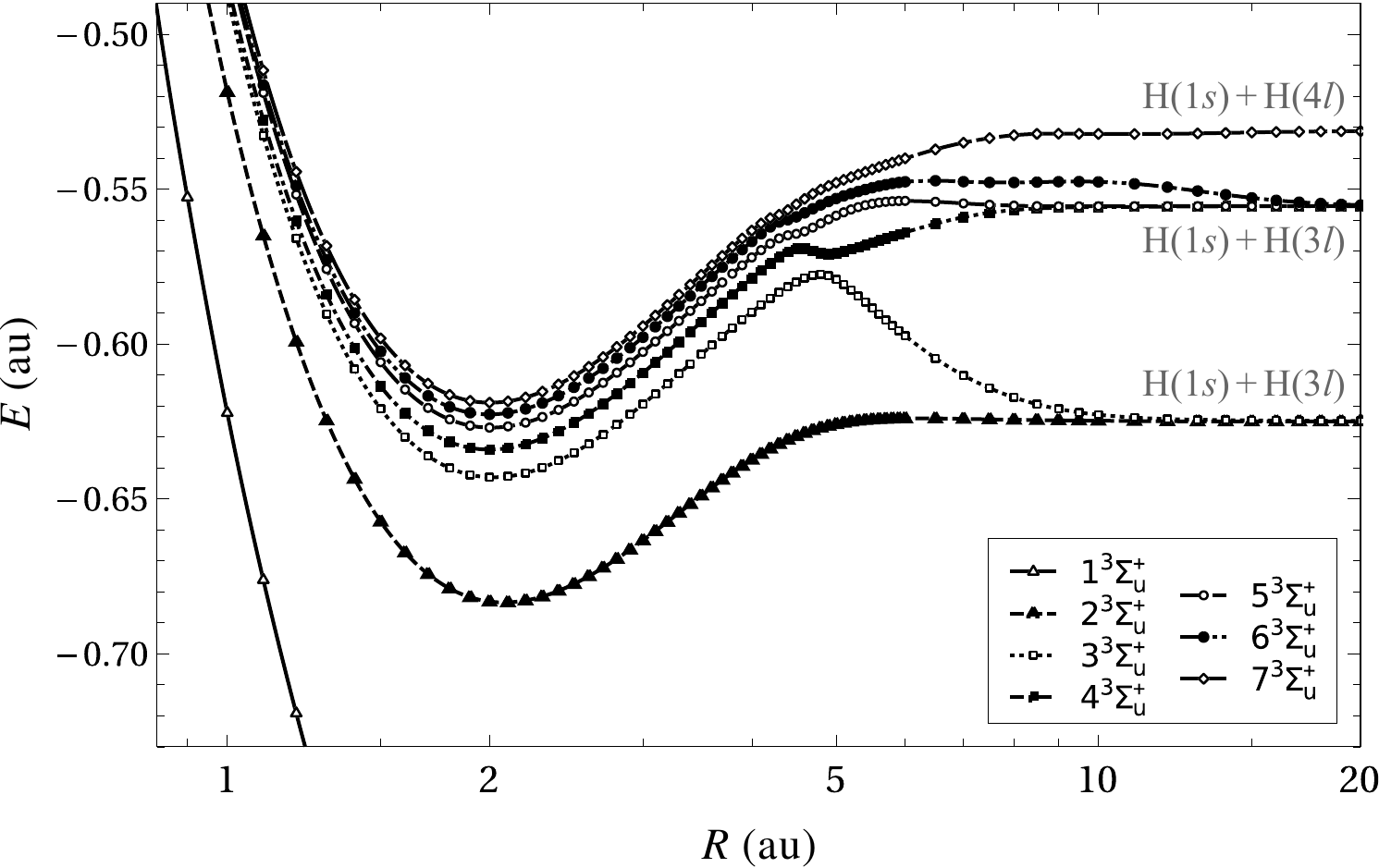}  
\caption{ BO potentials for $n^3\Sigma^{+}_u$ states, up to $n=7$, partial curve for $b\,1^3\Sigma^{+}_g$ from
Ref. \cite{gu} is shown for reference.}
\label{siut}
\end{figure}

\begin{table}[t]
	\centering
	\footnotesize
	\caption{Exemplary values of optimal nonlinear parameters for $n=7$ at various internuclear distances. Notice very small values of $y$,$u$ in sector 1 and
$x$,$w$ in sector 2, corresponding to very diffuse (high $n$) highly excited atomic orbitals.}
	\begin{tabular}{cc|cccc|cccc}
		\hline
		\hline
		\multirow{2}{*}{state} 		   &  \multirow{2}{*}{ $R$ }   & \multicolumn{4}{c|}{sector 1} &  \multicolumn{4}{c}{sector 2} \T \\
		\cline{3-10}
		     & & $ y$ & $x$ & $u$ & $w$ & $ y$ & $x$ & $u$ & $w$ \\
	\hline
		$7^1\Sigma^{+}_g$ & 2.0 &  0.000 & -0.423 & 0.097 & 0.821 & -0.080 & -0.122 & 0.637 & 0.262 \T \\
		$7^3\Sigma^{+}_g$ & 4.0 &  0.000 & -0.405 & 0.097 & 0.638 &  0.080 &  0.278 & 0.624 & 0.114 \\
		$7^1\Sigma^{+}_u$ & 6.0 & -0.061 & -0.514 & 0.146 & 0.619 & -0.285 &  0.080 & 0.578 & 0.080 \\
		$7^3\Sigma^{+}_u$ & 8.0 &  0.047 & -0.449 & 0.105 & 0.553 &  0.436 &  0.000 & 0.522 & 0.267 \\
	\hline
	\hline
\end{tabular}
\label{tab:params}
\end{table}
From inspection of potential curves in \Cref{sigs,sigt,sius,siut}, one evident universal feature for all the considered states
is the presence of principal minimum around 2 bohrs. Existence of a second minimum is mainly driven by the interaction of
energy levels with purely ionic H$^{+}$H$^{-}$ system. A modest drop in convergence rate can be noticed for the states:
$EF\,2^1\Sigma^{+}_g$, $H\bar{H}\,4^1\Sigma^{+}_g$, $7^1\Sigma^{+}_g$ and $B\,1^1\Sigma^{+}_u$,
$B''\bar{B}\,3^1\Sigma^{+}_u$, and $6^1\Sigma^{+}_u$ in regions of large $R$, where this interaction is most prevalent and configurations mix
markedly. This observation is consistent with the optimal values that nonlinear parameters attain, see \Cref{tab:ionic} for
exemplary values. For detailed analysis of ionic and covalent components we refer to Ref. \cite{corongiugs,corongiuus}.
In \Cref{tab:params} we present exemplary values of optimal nonlinear parameters for $n=7$ states. Evident asymmetry between
parameters of first and second electron justifies the use of more flexible KW basis over its more symmetric special cases.
We conjecture that the same basis type can be used as a remedy for the convergence rate of ionic states if our basis is augmented with a
third scale sector, improving the representation of electron correlation, when both are localized around the same nucleus,
similarly as it was performed in the calculations of HeH$^{+}$\cite{bohehp} BO potential and recognized as a feature of
\emph{in-out} electronic correlation in one-center, two-electron calculations of He \cite{drake}.

\begin{figure}[h]
\centering
\includegraphics[width=\columnwidth]{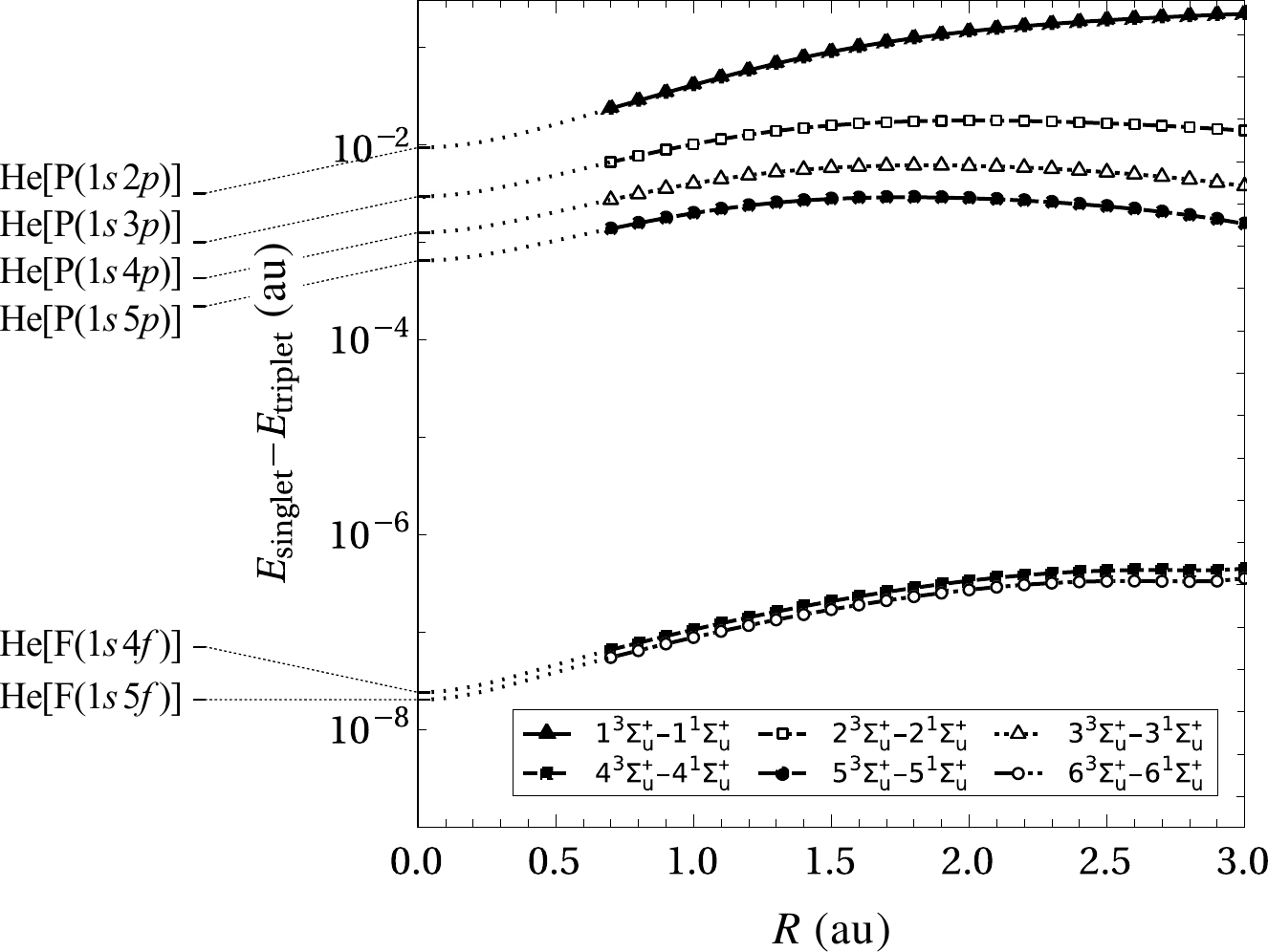}  
\caption{ Energy splittings between singlet and triplet $n\,\Sigma^{+}_u$ states at small internuclear
	distances. Labels on the $y$ axis indicate values of energy differences between helium $^1L(1snl)$ and $^3L(1snl)$
	states (here with $L=$P,F for the states under consideration).
	Dotted lines represent tentative assignments from calculated splittings to corresponding differences between
united atom ($R \rightarrow 0$) energies from Ref. \cite{drakehe}, which are in agreement with \cite{corongiuus,corongiut}.}
	
\label{heungerade}
\end{figure}

All the considered states dissociate as H$(1s)$ + H$(nl)$ with energies tending to the sum of its dissociation products
as $R \rightarrow \infty$, see \Cref{sigs,sigt,sius,siut}.
On the other hand, in the united atom limit, $R \rightarrow 0$, electronic configuration should approach those of He. 
Straightforward observation that the order of energies of atomic configurations He$[{}^{1,3}$L$(1snl)]$ is in general
different than corresponding H(1$s$) + H($nl$) dissociation products, together with well-known theorem that the BO
potential curves with the same symmetry  cannot cross, implies that the character of the states must change in
the region of intermediate $R$. The first qualitative approach aiming to predict such configuration mixing in terms of
so-called correlation diagrams can be attributed to Mulliken \cite{mulliken}, while
a detailed discussion on configuration mixing and matching of $\Sigma^{+}$ manifolds electronic densities to the corresponding atomic orbitals 
was presented in  the meticulous analysis by Corongiu and Clementi \cite{corongiugs,corongiuus,corongiut}. 
Here, we supplement this discussion  with the help of \Cref{heungerade}, where we present accurate splittings between singlet and triplet energies
approaching the united atom configuration, including ($1snf$) states which are split by values as small as $\sim 10^{-8}$ au.

\begin{table}[h]
	\centering
	\footnotesize
	\caption{Comparison of calculated BO energies (hartrees) for selected states at internuclear distance of 2 bohrs (vicinity of first minimum). Long dash indicates no data available for
	given method. Underlined digits present an improvement over previous most accurate result.}
	\setlength{\tabcolsep}{0.2em}
	\begin{tabular}{cclll}
		\hline
		\hline
		Method  & Ref.                                    & $EF\,2^1 \Sigma^{+}_g$                   & $H\bar{H}\,4^1\Sigma^{+}_u$	& $7\,{}^3\Sigma^{+}_u$ \T\B \\
\hline
		Full CI & \cite{corongiugs,corongiuus,corongiut}  & -0.717\,68                               & -0.634\,09                                & -0.618\,97  \T\\
		FC-LSE  & \cite{nakatsuji}                        & -0.717\,724\,7(96)                       & -0.634\,105(21)                           & -0.618\,207(64) \\
		GGTOs   & \cite{detmer}                           &  \text{---}                              & -0.634\,098\,15                           & \text{---}\\
		ECG     & \cite{komasaef}                        & -0.717\,715\,240                         & \text{---}   		                     & \text{---}\\
		KW      & \cite{orlikowski},\cite{staszewska2002} & -0.717\,715\,096                         & -0.634\,101\,622\,0                       & \text{---} \\
		KW      & this work                               & -0.717\,715\,2\underline{79\,148\,71}(3) & -0.634\,101\,6\underline{40\,058\,24}(1)                  & -0.618\,9\underline{69\,968\,931}(3) \\
	\hline
	\hline
\end{tabular}
\label{comparison}
\end{table}

For the sake of comparison with the most recent integral-free FC-LSE calculations by Nakashima and Nakatsuji \cite{nakatsuji}, 
we refer to the definition of H-square, see Eq. (24) of Ref. \cite{hsquare},
\begin{equation}
	\sigma^2_{FC-LSE} = \langle \Psi | \left(H-E\right)^2 | \Psi \rangle,
\end{equation}
which is utilized in Ref. \cite{nakatsuji} as a measure of wavefunction (and energy) accuracy.
For convenience, we define a relative ratio $p$ as,
\begin{equation}
	p = \frac{E_{var}-E_{FC-LSE}}{\sigma^2_{FC-LSE}},
	\label{eqp}
\end{equation}
where $E_{var}$ denotes the variational energy result as obtained in this work and $E_{FC-LSE}$ refers to results of the FC-LSE
method from Ref. \cite{nakatsuji}.
One observes, that  in the case of the highly excited $7^3\Sigma^{+}_u$ state,  $p$ is much greater than $1$ for the majority
of points and varies significantly (up to $\sim90$) along with internuclear distance with no visible trend. In contrast, another highly excited states available for the
reference, such as $6^1\Sigma^{+}_u$ or $7^1\Sigma^{+}_g$, have $p>1$ consistently in the region $R=2.0-6.0$ au.
For $n>4$ and all symmetries, systematically, more than half of the points fall outside the range of confidence for triplet
states, if H-square is given the interpretation of $1\sigma$ confidence interval. Therefore,
we conclude that H-square vastly underestimates the actual uncertainty of wavefunctions (and of energies) obtained in Ref.\,\cite{nakatsuji},
especially in case of triplet states and for large distances. Not surprisingly, a detailed comparison  in Table \ref{comparison}
of numerical results with their uncertainties for all the previous calculations for selected excited states at
given distance of  $R = 2.0$ au indicates that almost no previous calculations were able to correctly estimate numerical uncertainties. 






\section{Conclusions}
We have demonstrated that a relative accuracy of $10^{-10}$ or better can be reached for BO energies of all the excited
$n\,\Sigma^{+}$ states of H$_2$ up to $n=7$, with at most 18 000 basis functions of explicitly correlated exponential basis. 
With a two-sector unrestricted Kolos-Wolniewicz basis, a rapid, exponential convergence towards the CBS limit is attained, even
for exceptionally complicated electronic configurations of high singly-excited orbitals.

Adiabatic corrections are known to be of significant magnitude \cite{staszewska2002} and this work paves the way towards
their accurate evaluation with method similar to that presented in Ref. \cite{Ea}.
Ultimately, because the results obtained in this work are at least 6 orders of magnitude more accurate than any other
available in the literature, we believe that they will serve as a useful benchmark for less accurate computational methods.
Moreover, our  two-body two-center integrals with exponential functions can be used for any two-center systems, 
giving the possibility of achieving high-precision results for an arbitrary diatomic molecules.   

\section*{Supplementary Material}
For the complete list of our numerical results we refer to Tables S1--S6 for the BO energies of the $2\,^1\Sigma^{+}_g
- 7\,^1\Sigma^{+}_g$ states, Tables S7--S13 for the BO energies of the $1\,^3\Sigma^{+}_g - 7\,^3\Sigma^{+}_g$ states,
Tables S14--S20 for the BO energies of the $1\,^1\Sigma^{+}_u - 7\,^1\Sigma^{+}_u$ states, Tables S21--S26 for the BO energies of the $2\,^3\Sigma^{+}_u - 7\,^3\Sigma^{+}_u$ states in the Supplementary Material~\cite{supp}.

In addition, $dE/dR$, $-\langle \nabla_1^2 \rangle/2$ and $-\langle\nabla_1 \cdot \nabla_2\rangle/2$ along with energies are reported in raw text format of the Supplementary Material~\cite{supptxt}.

\section*{Acknowledgement}

M.S. acknowledges support from the National Science Center (Poland) under Grants No. 2020/36/T/ST2/00605 and No. 2017/27/B/ST2/02459.
K.P. acknowledges support from the National Science Center (Poland) under Grant No. 2017/27/B/ST2/02459.\\


\providecommand{\latin}[1]{#1}
\makeatletter
\providecommand{\doi}
  {\begingroup\let\do\@makeother\dospecials
  \catcode`\{=1 \catcode`\}=2 \doi@aux}
\providecommand{\doi@aux}[1]{\endgroup\texttt{#1}}
\makeatother
\providecommand*\mcitethebibliography{\thebibliography}
\csname @ifundefined\endcsname{endmcitethebibliography}
  {\let\endmcitethebibliography\endthebibliography}{}

\end{document}


{\centering
\section*{Supplementary Material}
\textbf{\large Accurate Born-Oppenheimer potentials for excited $\Sigma^{+}$ states of the hydrogen molecule}\\
\vspace*{0.5em}
Micha{\l} Si{\l}kowski, Magdalena Zientkiewicz, Krzysztof Pachucki \\

\textit{Faculty of Physics, University of Warsaw, Pasteura 5, 02-093 Warsaw, Poland} \\
(Dated: \today) \\
}
\vspace*{2.5em}

Contents: \\
\begin{itemize}
\item Tables S1 to S6: Born-Oppenheimer energies for $2\,^1\Sigma^{+}_g-7\,^1\Sigma^{+}_g$ states.\\
\item Tables S7 to S13: Born-Oppenheimer energies for $1\,^3\Sigma^{+}_g-7\,^3\Sigma^{+}_g$ states.\\
\item Tables S14 to S20: Born-Oppenheimer energies for $1\,^1\Sigma^{+}_u-7\,^1\Sigma^{+}_u$ states.\\
\item Tables S21 to S26: Born-Oppenheimer energies for $2\,^3\Sigma^{+}_u-7\,^3\Sigma^{+}_u$ states.\\
\end{itemize}

\vspace*{1.5em}
For additional quantities and operators, such as $dE/dR$, $-\langle \nabla_1^2 \rangle/2$ and $-\langle\nabla_1 \cdot \nabla_2\rangle/2$ along with $E$ itself, we refer to the plain text format of the Supplementary Material.

\begin{table}
	\centering
	\caption{Calculated BO energies of the $2^1\Sigma^{+}_g$ state in atomic units (hartree). Uncertainties originate purely from extrapolation to CBS limit.\\}
	\bgroup
	\setlength{\tabcolsep}{0.5em}
\begin{tabularx}{\textwidth}{@{}Yl|Yl@{}}
	\hline
	\hline
		\multicolumn{4}{c}{ \large $2^1\Sigma^{+}_g$} \T \T \B \B \\
	\hline
	\hline
		$R/au$ & $E/au$ & $R/au$ & $E/au$ \T \B \\
	\hline
		0.7 &   -0.316\,515\,768\,616\,4(4)   & 4.3 &   -0.714\,395\,560\,089\,51(2)  \\
0.8 &   -0.436\,425\,593\,131\,4(2)   & 4.4 &   -0.714\,520\,556\,251\,95(7)  \\
0.9 &   -0.520\,350\,468\,544\,3(2)   & 4.5 &   -0.714\,350\,459\,119\,94(9)  \\
1.0 &   -0.580\,085\,582\,389\,8(2)   & 4.6 &   -0.713\,919\,923\,180\,97(7)  \\
1.1 &   -0.623\,016\,874\,818\,74(8)  & 4.7 &   -0.713\,261\,360\,033\,09(7)  \\
1.2 &   -0.653\,971\,526\,522\,602(3) & 4.8 &   -0.712\,404\,663\,601\,50(9)  \\
\hline
	\hline
	\end{tabularx}
	\egroup
	\label{tab:t2sigs}
\end{table}

\begin{table}
	\centering
	\caption{Calculated BO energies of the $3^1\Sigma^{+}_g$ state in atomic units (hartree). Uncertainties originate purely from extrapolation to CBS limit.\\}
	\bgroup
	\setlength{\tabcolsep}{0.5em}
	\begin{tabularx}{\textwidth}{@{}Yl|Yl@{}}
	\hline
	\hline
		\multicolumn{4}{c}{ \large $3^1\Sigma^{+}_g$} \T \T \B \B \\
	\hline
	\hline
		$R/au$ & $E/au$ & $R/au$ & $E/au$ \T \B \\
	\hline
		0.7 &   -0.240\,521\,729\,67(3)       & 4.3 &   -0.641\,845\,665\,471\,78(9) \\
0.8 &   -0.361\,818\,601\,10(6)       & 4.4 &   -0.639\,999\,584\,864\,7(1)  \\
0.9 &   -0.447\,069\,355\,88(2)       & 4.5 &   -0.638\,295\,632\,428\,1(6)  \\
1.0 &   -0.508\,066\,130\,356(8)      & 4.6 &   -0.636\,732\,643\,288\,28(8) \\
1.1 &   -0.552\,582\,069\,001(2)      & 4.7 &   -0.635\,307\,374\,134\,00(7) \\
1.2 &   -0.585\,253\,000\,503\,92(8)  & 4.8 &   -0.634\,014\,937\,713\,46(9) \\
\hline
	\hline
	\end{tabularx}
	\egroup
	\label{tab:t3sigs}
\end{table}

\begin{table}
	\centering
	\caption{Calculated BO energies of the $4^1\Sigma^{+}_g$ state in atomic units (hartree). Uncertainties originate purely from extrapolation to CBS limit.\\}
	\bgroup
	\setlength{\tabcolsep}{0.5em}
	\begin{tabularx}{\textwidth}{@{}Yl|Yl@{}}
	\hline
	\hline
		\multicolumn{4}{c}{ \large $4^1\Sigma^{+}_g$} \T \T \B \B \\
	\hline
	\hline
		$R/au$ & $E/au$ & $R/au$ & $E/au$ \T \B \\
	\hline
		0.7 &   -0.238\,469\,687\,512\,3(7)     &  4.3 &   -0.599\,537\,485\,842\,4(4)              \\
0.8 &   -0.360\,391\,556\,326\,5(2)     &  4.4 &   -0.597\,731\,052\,126\,3(3)              \\
0.9 &   -0.446\,257\,979\,308\,4(3)     &  4.5 &   -0.596\,041\,711\,437\,1(3)              \\
1.0 &   -0.507\,859\,815\,261\,01(7)    &  4.6 &   -0.594\,468\,203\,035\,1(4)              \\
1.1 &   -0.552\,194\,141\,040\,3(6)     &  4.7 &   -0.593\,008\,641\,414\,4(4)              \\
1.2 &   -0.584\,281\,145\,968\,8(9)     &  4.8 &   -0.591\,660\,754\,131\,4(3)              \\
\hline
	\hline
	\end{tabularx}
	\egroup
	\label{tab:t4sigs}
\end{table}

\begin{table}
	\centering
	\caption{Calculated BO energies of the $5^1\Sigma^{+}_g$ state in atomic units (hartree). Uncertainties originate purely from extrapolation to CBS limit.\\}
	\bgroup
	\setlength{\tabcolsep}{0.5em}
	\begin{tabularx}{\textwidth}{@{}Yl|Yl@{}}
	\hline
	\hline
		\multicolumn{4}{c}{ \large $5^1\Sigma^{+}_g$} \T \T \B \B \\
	\hline
	\hline
		$R/au$ & $E/au$ & $R/au$ & $E/au$ \T \B \\
	\hline
		0.7 &   -0.214\,825\,011\,937(5)      & 4.3 &   -0.587\,453\,175\,649\,1(2)  \\
0.8 &   -0.336\,447\,866\,470(5)      & 4.4 &   -0.585\,261\,684\,172\,0(2)  \\
0.9 &   -0.422\,012\,812\,363(5)      & 4.5 &   -0.583\,192\,634\,594\,0(2)  \\
1.0 &   -0.483\,310\,990\,124(4)      & 4.6 &   -0.581\,243\,982\,248\,0(2)  \\
1.1 &   -0.527\,931\,910\,869\,4(7)   & 4.7 &   -0.579\,413\,005\,133\,2(2)  \\
1.2 &   -0.560\,542\,400\,786\,1(5)   & 4.8 &   -0.577\,696\,421\,046\,3(2)  \\
\hline
	\hline
	\end{tabularx}
	\egroup
	\label{tab:t5sigs}
\end{table}

\begin{table}
	\centering
	\caption{Calculated BO energies of the $6^1\Sigma^{+}_g$ state in atomic units (hartree). Uncertainties originate purely from extrapolation to CBS limit.\\}
	\bgroup
	\setlength{\tabcolsep}{0.5em}
	\begin{tabularx}{\textwidth}{@{}Yl|Yl@{}}
	\hline
	\hline
		\multicolumn{4}{c}{ \large $6^1\Sigma^{+}_g$} \T \T \B \B \\
	\hline
	\hline
		$R/au$ & $E/au$ & $R/au$ & $E/au$ \T \B \\
	\hline
		0.7 &   -0.214\,002\,325\,509\,2(9)   & 4.3 &   -0.571\,819\,715\,484(2)     \\
0.8 &   -0.335\,886\,648\,909(3)      & 4.4 &   -0.569\,694\,733\,503\,0(10) \\
0.9 &   -0.421\,710\,285\,628(2)      & 4.5 &   -0.567\,684\,437\,16(3)      \\
1.0 &   -0.483\,263\,818\,706\,8(6)   & 4.6 &   -0.565\,789\,362\,518(2)     \\
1.1 &   -0.527\,727\,162\,319(2)      & 4.7 &   -0.564\,010\,361\,903(2)     \\
1.2 &   -0.560\,088\,946\,762(5)      & 4.8 &   -0.562\,349\,067\,676(2)     \\
\hline
	\hline
	\end{tabularx}
	\egroup
	\label{tab:t6sigs}
\end{table}

\begin{table}
	\centering
	\caption{Calculated BO energies of the $7^1\Sigma^{+}_g$ state in atomic units (hartree). Uncertainties originate purely from extrapolation to CBS limit.\\}
	\bgroup
	\setlength{\tabcolsep}{0.5em}
	\begin{tabularx}{\textwidth}{@{}Yl|Yl@{}}
	\hline
	\hline
		\multicolumn{4}{c}{ \large $7^1\Sigma^{+}_g$} \T \T \B \B \\
	\hline
	\hline
		$R/au$ & $E/au$ & $R/au$ & $E/au$ \T \B \\
	\hline
		0.7 &   -0.203\,102\,622\,6(5)        &  4.3 &   -0.567\,038\,119\,622\,9(6)                               \\
0.8 &   -0.324\,839\,806\,2(2)        &  4.4 &   -0.564\,803\,850\,249\,7(5)                               \\
0.9 &   -0.410\,515\,404\,90(9)       &  4.5 &   -0.562\,691\,885\,771\,8(5)                               \\
1.0 &   -0.471\,920\,166\,180(3)      &  4.6 &   -0.560\,702\,734\,472\,9(4)                               \\
1.1 &   -0.516\,552\,689\,791(8)      &  4.7 &   -0.558\,837\,170\,226\,8(3)                               \\
1.2 &   -0.549\,140\,355\,703(4)      &  4.8 &   -0.557\,096\,631\,169\,3(4)                               \\
\hline
	\hline
	\end{tabularx}
	\egroup
	\label{tab:t7sigs}
\end{table}

\begin{table}
	\centering
	\caption{Calculated BO energies of the $1^3\Sigma^{+}_g$ state in atomic units (hartree). Uncertainties originate purely from extrapolation to CBS limit.\\}
	\bgroup
	\setlength{\tabcolsep}{0.5em}
	\begin{tabularx}{\textwidth}{@{}Yl|Yl@{}}
	\hline
	\hline
		\multicolumn{4}{c}{ \large $1^3\Sigma^{+}_g$} \T \T \B \B \\
	\hline
	\hline
		$R/au$ & $E/au$ & $R/au$ & $E/au$ \T \B \\
	\hline
		0.7 &  -0.342130928988667(6)     &  4.3   & -0.653195827155049(8) \\
0.8 &  -0.46143498347982(4)      &  4.4   & -0.65073706182336(2)  \\
0.9 &  -0.544767040091894(3)     &  4.5   & -0.6484429623517(4)   \\
1.0 &  -0.60392224803181505(4) &    4.6   & -0.6463109963342(4)   \\
1.1 &  -0.6462856755609065(2)  &    4.7   & -0.644337697996883(3) \\
1.2 &  -0.67668314180341571(6) &    4.8   & -0.6425186973868(9)   \\
\hline
	\hline
	\end{tabularx}
	\egroup
	\label{tab:t1sigt}
\end{table}

\begin{table}
	\centering
	\caption{Calculated BO energies of the $2^3\Sigma^{+}_g$ state in atomic units (hartree). Uncertainties originate purely from extrapolation to CBS limit.\\}
	\bgroup
	\setlength{\tabcolsep}{0.5em}
	\begin{tabularx}{\textwidth}{@{}Yl|Yl@{}}
	\hline
	\hline
		\multicolumn{4}{c}{ \large $2^3\Sigma^{+}_g$} \T \T \B \B \\
	\hline
	\hline
		$R/au$ & $E/au$ & $R/au$ & $E/au$ \T \B \\
	\hline
		0.7 & -0.247\,264\,442\,604\,2(4)   &   4.3 & -0.614\,697\,062\,021\,16(2)     \\ 
0.8 & -0.368\,441\,813\,489\,63(8)  &   4.4 & -0.613\,958\,695\,445\,20(2)     \\ 
0.9 & -0.453\,573\,150\,521\,41(2)  &   4.5 & -0.613\,364\,273\,414\,49(2)     \\ 
1.0 & -0.514\,450\,615\,877\,05(2)  &   4.6 & -0.612\,901\,703\,757\,55(2)     \\ 
1.1 & -0.558\,459\,323\,262\,524(3) &   4.7 & -0.612\,558\,617\,305\,02(2)     \\ 
1.2 & -0.590\,426\,867\,896\,656(9) &   4.8 & -0.612\,322\,805\,243\,64(3)     \\
\hline
	\hline
	\end{tabularx}
	\egroup
	\label{tab:t2sigt}
\end{table}

\begin{table}
	\centering
	\caption{Calculated BO energies of the $3^3\Sigma^{+}_g$ state in atomic units (hartree). Uncertainties originate purely from extrapolation to CBS limit.\\}
	\bgroup
	\setlength{\tabcolsep}{0.5em}
	\begin{tabularx}{\textwidth}{@{}Yl|Yl@{}}
	\hline
	\hline
		\multicolumn{4}{c}{ \large $3^3\Sigma^{+}_g$} \T \T \B \B \\
	\hline
	\hline
		$R/au$ & $E/au$ & $R/au$ & $E/au$ \T \B \\
	\hline
		0.7 & -0.238\,502\,584\,901\,1(5)      & 4.3 & -0.589\,777\,941\,367\,421\,9(2) \\
0.8 & -0.360\,428\,580\,812\,6(3)      & 4.4 & -0.587\,397\,930\,925\,423\,8(4) \\
0.9 & -0.446\,299\,293\,025\,778(8)    & 4.5 & -0.585\,147\,011\,234\,322\,6(3) \\
1.0 & -0.507\,905\,409\,337\,34(3)     & 4.6 & -0.583\,024\,545\,273\,709\,1(2) \\
1.1 & -0.552\,631\,673\,439\,913(6)    & 4.7 & -0.581\,029\,432\,141\,636\,3(2) \\
1.2 & -0.585\,305\,949\,989\,59(2)     & 4.8 & -0.579\,160\,144\,174\,822\,4(2) \\
\hline
	\hline
	\end{tabularx}
	\egroup
	\label{tab:t3sigt}
\end{table}

\begin{table}
	\centering
	\caption{Calculated BO energies of the $4^3\Sigma^{+}_g$ state in atomic units (hartree). Uncertainties originate purely from extrapolation to CBS limit.\\}
	\bgroup
	\setlength{\tabcolsep}{0.5em}
	\begin{tabularx}{\textwidth}{@{}Yl|Yl@{}}
	\hline
	\hline
		\multicolumn{4}{c}{ \large $4^3\Sigma^{+}_g$} \T \T \B \B \\
	\hline
	\hline
		$R/au$ & $E/au$ & $R/au$ & $E/au$ \T \B \\
	\hline
		0.7 & -0.217\,536\,646\,295\,9(7)      & 4.3 & -0.577\,420\,960\,308\,37(7)     \\
0.8 & -0.339\,120\,034\,158\,1(5)      & 4.4 & -0.575\,545\,117\,964\,44(6)     \\
0.9 & -0.424\,644\,979\,746\,7(2)      & 4.5 & -0.573\,775\,776\,542\,70(4)     \\
1.0 & -0.485\,902\,620\,977\,2(2)      & 4.6 & -0.572\,111\,684\,791\,37(6)     \\
1.1 & -0.530\,277\,685\,752\,4(2)      & 4.7 & -0.570\,552\,232\,831\,47(6)     \\
1.2 & -0.562\,597\,748\,380\,5(3)      & 4.8 & -0.569\,097\,537\,363\,06(3)     \\
\hline
	\hline
	\end{tabularx}
	\egroup
	\label{tab:t4sigt}
\end{table}

\begin{table}
	\centering
	\caption{Calculated BO energies of the $5^3\Sigma^{+}_g$ state in atomic units (hartree). Uncertainties originate purely from extrapolation to CBS limit.\\}
	\bgroup
	\setlength{\tabcolsep}{0.5em}
	\begin{tabularx}{\textwidth}{@{}Yl|Yl@{}}
	\hline
	\hline
		\multicolumn{4}{c}{ \large $5^3\Sigma^{+}_g$} \T \T \B \B \\
	\hline
	\hline
		$R/au$ & $E/au$ & $R/au$ & $E/au$ \T \B \\
	\hline
		0.7 & -0.214\,020\,967\,24(4)       & 4.3 & -0.567\,803\,978\,955\,034(4)\\
0.8 & -0.335\,907\,516\,563(4)      & 4.4 & -0.565\,473\,005\,504\,785(3)\\
0.9 & -0.421\,733\,421\,4735(7)     & 4.5 & -0.563\,266\,868\,994\,272(2)\\
1.0 & -0.483\,289\,150\,953(2)      & 4.6 & -0.561\,188\,647\,017\,841(3)\\
1.1 & -0.527\,959\,195\,692(4)      & 4.7 & -0.559\,243\,775\,865\,617(5)\\
1.2 & -0.560\,571\,136\,081\,5(2)   & 4.8 & -0.557\,442\,655\,718\,874(5)\\
\hline
	\hline
	\end{tabularx}
	\egroup
	\label{tab:t5sigt}
\end{table}

\begin{table}
	\centering
	\caption{Calculated BO energies of the $6^3\Sigma^{+}_g$ state in atomic units (hartree). Uncertainties originate purely from extrapolation to CBS limit.\\}
	\bgroup
	\setlength{\tabcolsep}{0.5em}
	\begin{tabularx}{\textwidth}{@{}Yl|Yl@{}}
	\hline
	\hline
		\multicolumn{4}{c}{ \large $6^3\Sigma^{+}_g$} \T \T \B \B \\
	\hline
	\hline
		$R/au$ & $E/au$ & $R/au$ & $E/au$ \T \B \\
	\hline
		0.7 & -0.204\,455\,278\,331(6)     & 4.3 & -0.562\,209\,395\,510\,0(3)  \\
0.8 & -0.326\,175\,437\,534\,4(3)  & 4.4 & -0.560\,082\,924\,728\,9(2)  \\
0.9 & -0.411\,833\,594\,334\,1(6)  & 4.5 & -0.558\,063\,138\,519\,9(2)  \\
1.0 & -0.473\,220\,487\,327\,9(4)  & 4.6 & -0.556\,150\,678\,784\,5(2)  \\
1.1 & -0.517\,720\,655\,208\,4(5)  & 4.7 & -0.554\,348\,320\,135\,1(2)  \\
1.2 & -0.550\,161\,602\,178\,3(3)  & 4.8 & -0.552\,664\,220\,177\,4(2)  \\
\hline
	\hline
	\end{tabularx}
	\egroup
	\label{tab:t6sigt}
\end{table}

\begin{table}
	\centering
	\caption{Calculated BO energies of the $7^3\Sigma^{+}_g$ state in atomic units (hartree). Uncertainties originate purely from extrapolation to CBS limit.\\}
	\bgroup
	\setlength{\tabcolsep}{0.5em}
	\begin{tabularx}{\textwidth}{@{}Yl|Yl@{}}
	\hline
	\hline
		\multicolumn{4}{c}{ \large $7^3\Sigma^{+}_g$} \T \T \B \B \\
	\hline
	\hline
		$R/au$ & $E/au$ & $R/au$ & $E/au$ \T \B \\
	\hline
		0.7 & -0.202\,702\,053\,42(5)       & 4.3 & -0.558\,752\,897\,983\,14(5) \\
0.8 & -0.324\,573\,662\,75(3)       & 4.4 & -0.556\,456\,501\,450\,1(2)  \\
0.9 & -0.410\,382\,606\,17(2)       & 4.5 & -0.554\,270\,168\,415\,8(7)  \\
1.0 & -0.471\,919\,268\,627(8)      & 4.6 & -0.552\,192\,994\,565\,54(9) \\
1.1 & -0.516\,568\,046\,016(5)      & 4.7 & -0.550\,224\,011\,251\,8(4)  \\
1.2 & -0.549\,156\,416\,485(3)      & 4.8 & -0.548\,364\,200\,864\,39(2) \\
\hline
	\hline
	\end{tabularx}
	\egroup
	\label{tab:t7sigt}
\end{table}

\begin{table}
	\centering
	\caption{Calculated BO energies of the $1^1\Sigma^{+}_u$ state in atomic units (hartree). Uncertainties originate purely from extrapolation to CBS limit.\\}
	\bgroup
	\setlength{\tabcolsep}{0.5em}
	\begin{tabularx}{\textwidth}{@{}Yl|Yl@{}}
	\hline
	\hline
		\multicolumn{4}{c}{ \large $1^1\Sigma^{+}_u$} \T \T \B \B \\
	\hline
	\hline
		$R/au$ & $E/au$ & $R/au$ & $E/au$ \T \B \\
	\hline
		0.7 &  -0.342130928988667(6)     &  4.3   & -0.653195827155049(8) \\
0.8 &  -0.46143498347982(4)      &  4.4   & -0.65073706182336(2)  \\
0.9 &  -0.544767040091894(3)     &  4.5   & -0.6484429623517(4)   \\
1.0 &  -0.60392224803181505(4) &    4.6   & -0.6463109963342(4)   \\
1.1 &  -0.6462856755609065(2)  &    4.7   & -0.644337697996883(3) \\
1.2 &  -0.67668314180341571(6) &    4.8   & -0.6425186973868(9)   \\
\hline
	\hline
	\end{tabularx}
	\egroup
	\label{tab:t1sius}
\end{table}

\begin{table}
	\centering
	\caption{Calculated BO energies of the $2^1\Sigma^{+}_u$ state in atomic units (hartree). Uncertainties originate purely from extrapolation to CBS limit.\\}
	\bgroup
	\setlength{\tabcolsep}{0.5em}
	\begin{tabularx}{\textwidth}{@{}Yl|Yl@{}}
	\hline
	\hline
		\multicolumn{4}{c}{ \large $2^1\Sigma^{+}_u$} \T \T \B \B \\
	\hline
	\hline
		$R/au$ & $E/au$ & $R/au$ & $E/au$ \T \B \\
	\hline
		0.7 & -0.238\,540\,424\,455(2)      & 4.3 & -0.628\,854\,772\,172\,40(9)  \\
0.8 & -0.360\,658\,634\,698(8)      & 4.4 & -0.628\,289\,641\,463\,10(7)  \\
0.9 & -0.446\,755\,626\,041(9)      & 4.5 & -0.627\,830\,572\,168\,30(7)  \\
1.0 & -0.508\,623\,142\,870(2)      & 4.6 & -0.627\,462\,667\,471\,31(7)  \\
1.1 & -0.553\,646\,688\,943\,9(3)   & 4.7 & -0.627\,171\,731\,657\,94(6)  \\
1.2 & -0.586\,654\,492\,022\,5(5)   & 4.8 & -0.626\,944\,690\,182\,90(6)  \\
\hline
	\hline
	\end{tabularx}
	\egroup
	\label{tab:t2sius}
\end{table}

\begin{table}
	\centering
	\caption{Calculated BO energies of the $3^1\Sigma^{+}_u$ state in atomic units (hartree). Uncertainties originate purely from extrapolation to CBS limit.\\}
	\bgroup
	\setlength{\tabcolsep}{0.5em}
	\begin{tabularx}{\textwidth}{@{}Yl|Yl@{}}
	\hline
	\hline
		\multicolumn{4}{c}{ \large $3^1\Sigma^{+}_u$} \T \T \B \B \\
	\hline
	\hline
		$R/au$ & $E/au$ & $R/au$ & $E/au$ \T \B \\
	\hline
		0.7 & -0.214\,021\,007\,253(4)      & 4.3 & -0.584\,590\,871\,106\,42\\
0.8 & -0.335\,988\,249\,501(4)      & 4.4 & -0.582\,905\,438\,381\,35\\
0.9 & -0.421\,909\,252\,39(3)       & 4.5 & -0.581\,291\,512\,643\,71\\
1.0 & -0.483\,574\,887\,235(4)      & 4.6 & -0.579\,743\,429\,521\,85\\
1.1 & -0.528\,369\,903\,241(3)      & 4.7 & -0.578\,257\,136\,745\,01\\
1.2 & -0.561\,121\,954\,872(3)      & 4.8 & -0.576\,830\,034\,039\,14\\
\hline
	\hline
	\end{tabularx}
	\egroup
	\label{tab:t3sius}
\end{table}

\begin{table}
	\centering
	\caption{Calculated BO energies of the $4^1\Sigma^{+}_u$ state in atomic units (hartree). Uncertainties originate purely from extrapolation to CBS limit.\\}
	\bgroup
	\setlength{\tabcolsep}{0.5em}
	\begin{tabularx}{\textwidth}{@{}Yl|Yl@{}}
	\hline
	\hline
		\multicolumn{4}{c}{ \large $4^1\Sigma^{+}_u$} \T \T \B \B \\
	\hline
	\hline
		$R/au$ & $E/au$ & $R/au$ & $E/au$ \T \B \\
	\hline
		0.7 & -0.213\,906\,164\,232\,6(4)   & 4.3 & -0.571\,691\,450\,940\,6(3)  \\
0.8 & -0.335\,769\,148\,880\,7(4)   & 4.4 & -0.569\,647\,202\,397\,9(2)  \\
0.9 & -0.421\,568\,270\,896\,8(8)   & 4.5 & -0.567\,776\,268\,871\,9(3)  \\
1.0 & -0.483\,093\,840\,186\,66(4)  & 4.6 & -0.566\,101\,351\,451\,7(4)  \\
1.1 & -0.527\,730\,172\,074\,9(2)   & 4.7 & -0.564\,654\,775\,476\,3(7)  \\
1.2 & -0.560\,304\,664\,538\,7(4)   & 4.8 & -0.563\,480\,505\,873\,2(2)  \\
\hline
	\hline
	\end{tabularx}
	\egroup
	\label{tab:t4sius}
\end{table}

\begin{table}
	\centering
	\caption{Calculated BO energies of the $5^1\Sigma^{+}_u$ state in atomic units (hartree). Uncertainties originate purely from extrapolation to CBS limit.\\}
	\bgroup
	\setlength{\tabcolsep}{0.5em}
	\begin{tabularx}{\textwidth}{@{}Yl|Yl@{}}
	\hline
	\hline
		\multicolumn{4}{c}{ \large $5^1\Sigma^{+}_u$} \T \T \B \B \\
	\hline
	\hline
		$R/au$ & $E/au$ & $R/au$ & $E/au$ \T \B \\
	\hline
		0.7 & -0.202\,698\,747\,624(7)          & 4.3 & -0.566\,204\,349\,790(2)          \\
0.8 & -0.324\,611\,637\,060(9)          & 4.4 & -0.564\,235\,019\,697\,9(5)       \\
0.9 & -0.410\,469\,233\,105(9)          & 4.5 & -0.562\,363\,520\,035\,3(6)       \\
1.0 & -0.472\,062\,106\,988(7)          & 4.6 & -0.560\,591\,206\,703\,2(10)      \\
1.1 & -0.516\,774\,760\,61(4)           & 4.7 & -0.558\,921\,660\,050(2)          \\
1.2 & -0.549\,434\,677\,53(2)           & 4.8 & -0.557\,361\,064\,460\,6(9)       \\
\hline
	\hline
	\end{tabularx}
	\egroup
	\label{tab:t5sius}
\end{table}

\begin{table}
	\centering
	\caption{Calculated BO energies of the $6^1\Sigma^{+}_u$ state in atomic units (hartree). Uncertainties originate purely from extrapolation to CBS limit.\\}
	\bgroup
	\setlength{\tabcolsep}{0.5em}
	\begin{tabularx}{\textwidth}{@{}Yl|Yl@{}}
	\hline
	\hline
		\multicolumn{4}{c}{ \large $6^1\Sigma^{+}_u$} \T \T \B \B \\
	\hline
	\hline
		$R/au$ & $E/au$ & $R/au$ & $E/au$ \T \B \\
	\hline
		0.7 & -0.202\,641\,543\,808(4)     & 4.3 & -0.559\,677\,367\,918(9) \\
0.8 & -0.324\,500\,857\,535(2)     & 4.4 & -0.557\,572\,619\,296(1) \\
0.9 & -0.410\,295\,833\,993(2)     & 4.5 & -0.555\,648\,421\,851(2) \\
1.0 & -0.471\,816\,777\,429\,8(2)  & 4.6 & -0.553\,939\,738\,997(3) \\
1.1 & -0.516\,447\,995\,872\,22(3) & 4.7 & -0.552\,494\,223\,188(5) \\
1.2 & -0.549\,016\,878\,249\,94(1) & 4.8 & -0.551\,354\,534\,192(2) \\
\hline
	\hline
	\end{tabularx}
	\egroup
	\label{tab:t6sius}
\end{table}

\begin{table}
	\centering
	\caption{Calculated BO energies of the $7^1\Sigma^{+}_u$ state in atomic units (hartree). Uncertainties originate purely from extrapolation to CBS limit.\\}
	\bgroup
	\setlength{\tabcolsep}{0.5em}
	\begin{tabularx}{\textwidth}{@{}Yl|Yl@{}}
	\hline
	\hline
		\multicolumn{4}{c}{ \large $7^1\Sigma^{+}_u$} \T \T \B \B \\
	\hline
	\hline
		$R/au$ & $E/au$ & $R/au$ & $E/au$ \T \B \\
	\hline
		0.7 & -0.196\,556\,267\,4(4)       & 4.3 & -0.556\,938\,194\,00(3)      \\
0.8 & -0.318\,444\,971\,68(4)      & 4.4 & -0.554\,868\,900\,09(5)      \\
0.9 & -0.404\,274\,355\,18(4)      & 4.5 & -0.552\,924\,316\,40(7)      \\
1.0 & -0.465\,834\,863\,6(1)       & 4.6 & -0.551\,116\,724\,31(9)      \\
1.1 & -0.510\,510\,900\,51(4)      & 4.7 & -0.549\,464\,235\,12(9)       \\
1.2 & -0.543\,129\,888\,44(3)      & 4.8 & -0.547\,989\,623\,6(2)        \\
\hline
	\hline
	\end{tabularx}
	\egroup
	\label{tab:t7sius}
\end{table}

\begin{table}
	\centering
	\caption{Calculated BO energies of the $2^3\Sigma^{+}_u$ state in atomic units (hartree). Uncertainties originate purely from extrapolation to CBS limit.\\}
	\bgroup
	\setlength{\tabcolsep}{0.5em}
	\begin{tabularx}{\textwidth}{@{}Yl|Yl@{}}
	\hline
	\hline
		\multicolumn{4}{c}{ \large $2^3\Sigma^{+}_u$} \T \T \B \B \\
	\hline
	\hline
		$R/au$ & $E/au$ & $R/au$ & $E/au$ \T \B \\
	\hline
		0.7 & -0.245\,149\,517\,334\,5(2)  & 4.3 & -0.632\,300\,341\,455\,545(2) \\  
0.8 & -0.368\,332\,096\,096\,89(4) & 4.4 & -0.630\,946\,270\,507\,275(2) \\  
0.9 & -0.455\,605\,952\,653\,3(2)  & 4.5 & -0.629\,752\,332\,465\,679(3) \\  
1.0 & -0.518\,727\,298\,884\,55(2) & 4.6 & -0.628\,711\,272\,602\,945(2) \\  
1.1 & -0.565\,032\,494\,492\,68(4) & 4.7 & -0.627\,813\,848\,138\,214(2) \\  
1.2 & -0.599\,293\,589\,547\,7(1)  & 4.8 & -0.627\,049\,230\,707\,203(2) \\
\hline
	\hline
	\end{tabularx}
	\egroup
	\label{tab:t2siut}
\end{table}

\begin{table}
	\centering
	\caption{Calculated BO energies of the $3^3\Sigma^{+}_u$ state in atomic units (hartree). Uncertainties originate purely from extrapolation to CBS limit.\\}
	\bgroup
	\setlength{\tabcolsep}{0.5em}
	\begin{tabularx}{\textwidth}{@{}Yl|Yl@{}}
	\hline
	\hline
		\multicolumn{4}{c}{ \large $3^3\Sigma^{+}_u$} \T \T \B \B \\
	\hline
	\hline
		$R/au$ & $E/au$ & $R/au$ & $E/au$ \T \B \\
	\hline
		0.7 & -0.216\,737\,782\,218(7)                  & 4.3 & -0.583\,343\,745\,660\,32(2)      \\
0.8 & -0.339\,106\,607\,001(8)                  & 4.4 & -0.581\,570\,068\,610\,58(2)      \\
0.9 & -0.425\,460\,109\,053(8)                  & 4.5 & -0.580\,015\,682\,532\,28(2)      \\
1.0 & -0.487\,572\,969\,892\,3(2)               & 4.6 & -0.578\,731\,807\,044\,21(2)        \\
1.1 & -0.532\,810\,065\,278\,4(4)               & 4.7 & -0.577\,827\,919\,057\,791(10)      \\
1.2 & -0.565\,978\,385\,627\,6(3)               & 4.8 & -0.577\,514\,960\,007\,280(8)       \\
\hline
	\hline
	\end{tabularx}
	\egroup
	\label{tab:t3siut}
\end{table}

\begin{table}
	\centering
	\caption{Calculated BO energies of the $4^3\Sigma^{+}_u$ state in atomic units (hartree). Uncertainties originate purely from extrapolation to CBS limit.\\}
	\bgroup
	\setlength{\tabcolsep}{0.5em}
	\begin{tabularx}{\textwidth}{@{}Yl|Yl@{}}
	\hline
	\hline
		\multicolumn{4}{c}{ \large $4^3\Sigma^{+}_u$} \T \T \B \B \\
	\hline
	\hline
		$R/au$ & $E/au$ & $R/au$ & $E/au$ \T \B \\
	\hline
		0.7 & -0.213\,906\,229\,898\,5(4)  & 4.3 & -0.571\,961\,453\,137\,25(2)   \\
0.8 & -0.335\,769\,226\,266\,6(3)  & 4.4 & -0.570\,256\,295\,832\,041(5)  \\
0.9 & -0.421\,568\,361\,683\,52(6) & 4.5 & -0.569\,255\,219\,981\,32(2)   \\
1.0 & -0.483\,093\,946\,080\,72(9) & 4.6 & -0.569\,318\,920\,199\,38(3)   \\
1.1 & -0.527\,730\,294\,814\,35(9) & 4.7 & -0.570\,052\,220\,598\,22(5)   \\
1.2 & -0.560\,304\,805\,869\,79(6) & 4.8 & -0.570\,759\,342\,615\,79(7)   \\
\hline
	\hline
	\end{tabularx}
	\egroup
	\label{tab:t4siut}
\end{table}

\begin{table}
	\centering
	\caption{Calculated BO energies of the $5^3\Sigma^{+}_u$ state in atomic units (hartree). Uncertainties originate purely from extrapolation to CBS limit.\\}
	\bgroup
	\setlength{\tabcolsep}{0.5em}
	\begin{tabularx}{\textwidth}{@{}Yl|Yl@{}}
	\hline
	\hline
		\multicolumn{4}{c}{ \large $5^3\Sigma^{+}_u$} \T \T \B \B \\
	\hline
	\hline
		$R/au$ & $E/au$ & $R/au$ & $E/au$ \T \B \\
	\hline
		0.7 & -0.204\,068\,665\,129(4)                  & 4.3 & -0.565\,957\,163\,030\,17(3)     \\
0.8 & -0.326\,175\,387\,839(2)                  & 4.4 & -0.564\,959\,311\,005\,42(6)     \\
0.9 & -0.412\,239\,095\,498(4)                  & 4.5 & -0.564\,486\,638\,900\,60(5)     \\
1.0 & -0.474\,042\,043\,214(7)                  & 4.6 & -0.563\,696\,078\,158\,71(3)     \\
1.1 & -0.518\,958\,982\,599(5)                  & 4.7 & -0.562\,470\,949\,532\,53(3)     \\
1.2 & -0.551\,807\,651\,512(5)                  & 4.8 & -0.561\,123\,149\,558\,22(5)     \\
\hline
	\hline
	\end{tabularx}
	\egroup
	\label{tab:t5siut}
\end{table}

\begin{table}
	\centering
	\caption{Calculated BO energies of the $6^3\Sigma^{+}_u$ state in atomic units (hartree). Uncertainties originate purely from extrapolation to CBS limit.\\}
	\bgroup
	\setlength{\tabcolsep}{0.5em}
	\begin{tabularx}{\textwidth}{@{}Yl|Yl@{}}
	\hline
	\hline
		\multicolumn{4}{c}{ \large $6^3\Sigma^{+}_u$} \T \T \B \B \\
	\hline
	\hline
		$R/au$ & $E/au$ & $R/au$ & $E/au$ \T \B \\
	\hline
		0.7 & -0.202\,641\,598\,69(2)                  & 4.3 & -0.560\,674\,017\,760\,1(3)  \\
0.8 & -0.324\,500\,922\,089(5)                 & 4.4 & -0.559\,945\,453\,737\,23(2) \\
0.9 & -0.410\,295\,909\,591\,1(3)              & 4.5 & -0.559\,001\,202\,875\,92(3)    \\
1.0 & -0.471\,816\,865\,441\,68(2)             & 4.6 & -0.557\,725\,475\,834\,69(5)     \\
1.1 & -0.516\,448\,097\,681\,65(4)             & 4.7 & -0.556\,395\,851\,361\,61(8)     \\
1.2 & -0.549\,016\,995\,231\,0(3)              & 4.8 & -0.555\,152\,589\,518\,86(3)    \\
\hline
	\hline
	\end{tabularx}
	\egroup
	\label{tab:t6siut}
\end{table}

\begin{table}
	\centering
	\caption{Calculated BO energies of the $7^3\Sigma^{+}_u$ state in atomic units (hartree). Uncertainties originate purely from extrapolation to CBS limit.\\}
	\bgroup
	\setlength{\tabcolsep}{0.5em}
	\begin{tabularx}{\textwidth}{@{}Yl|Yl@{}}
	\hline
	\hline
		\multicolumn{4}{c}{ \large $7^3\Sigma^{+}_u$} \T \T \B \B \\
	\hline
	\hline
		$R/au$ & $E/au$ & $R/au$ & $E/au$ \T \B \\
	\hline
		0.7 & -0.197\,341\,162\,11(6)                  & 4.3 & -0.558\,292\,990\,692(2)     \\
0.8 & -0.319\,338\,039\,10(5)                  & 4.4 & -0.556\,768\,762\,632\,0(9)  \\
0.9 & -0.405\,281\,585\,88(3)                  & 4.5 & -0.554\,952\,260\,223(2)     \\
1.0 & -0.466\,957\,462\,89(4)                  & 4.6 & -0.553\,189\,345\,326\,9(10) \\
1.1 & -0.511\,744\,599\,48(3)                  & 4.7 & -0.551\,586\,044\,062(4)     \\
1.2 & -0.544\,465\,066\,29(3)                  & 4.8 & -0.550\,190\,627\,456(5)     \\
\hline
	\hline
	\end{tabularx}
	\egroup
	\label{tab:t7siut}
\end{table}